\documentclass{PoS}

\title{Core-collapse Supernovae as Cosmic Ray Sources}

\ShortTitle{Supernovae as Cosmic Ray Sources}

\author{\speaker{Vikram Dwarkadas}\\
       University of Chicago\\
        E-mail: \email{vikram@astro.uchicago.edu}}

\author{Alexandre Marcowith\\
        Universite Montpellier\\
        E-mail: \email{almarcowith@gmail.com}}

\author{Matthieu Renaud\\
        Universite Montpellier\\
        E-mail: \email{matthieu.renaud@umontpellier.fr}}

\author{Vincent Tatischeff\\
        Universite Paris-Sud\\
        E-mail: \email{Vincent.Tatischeff@csnsm.in2p3.fr}}

\author{Gwenael Giacinti\\
        Max-Planck-Institut fur Kernphysik\\
        E-mail: \email{Gwenael.Giacinti@mpi-hd.mpg.de}}

\abstract{Core-collapse supernovae produce fast shocks which expand
  into the dense circumstellar medium (CSM) of the stellar
  progenitor. Cosmic rays (CRs) accelerated at these shocks can induce
  the growth of electromagnetic fluctuations in the pre-shock
  medium. Using a self-similar description for the shock evolution, we
  calculate the growth time-scales of CR driven instabilities for SNe
  in general, and SN 1993J in particular.  We find that extended SN
  shocks can trigger fast intra-day instabilities, strong magnetic
  field amplification, and CR acceleration. In particular, the
  non-resonant streaming instability can contribute to about 50 per
  cent of the magnetic field intensity deduced from radio data. This
  results in the acceleration of CR particles to energies of 1-10 PeV
  within a few days after the shock breakout. }

\FullConference{36th International Cosmic Ray Conference -ICRC2019-\\
		July 24th - August 1st, 2019\\
		Madison, WI, U.S.A.}

\def\gtrsim{\mathrel{\hbox{\rlap{\hbox{\lower4pt\hbox{$\sim$}}}\hbox{$>$}}}}

\newcommand\be {\begin{equation}}
\newcommand\en{\end{equation}}

\newcommand{\noi}{\noindent}
\newcommand{\ee}{\end{equation}}

\newcommand{\beq}{\begin{equation}}
\newcommand{\eeq}{\end{equation}}
\newcommand{\bea}{\begin{eqnarray}}
\newcommand{\eea}{\end{eqnarray}}
\newcommand{\bei}{\begin{itemize}}
\newcommand{\eei}{\end{itemize}}
\newcommand{\bee}{\begin{enumerate}}
\newcommand{\eee}{\end{enumerate}}

\begin{document}

\section{Introduction}
High-energy cosmic rays (CRs) are likely accelerated in fast shocks
produced in very energetic events \cite{Bell78}. CRs above an energy
of $10^{17}-10^{18}$ eV are expected to arise from extragalactic
sources. Below this energy the sources are thought to be Galactic,
such as young supernova remnants (SNRs). Many of these SNRs have
magnetic field strengths much larger than could be expected from shock
compression of the interstellar medium magnetic field.

The process of amplification of the magnetic field is unclear. One
argument is that magnetic field amplification (MFA) originates from
plasma instabilities driven by CR ions \cite{Bell04, Pelletier06,
  Marcowith06, Zira08, Amato09, Bykov11}.

An important argument raised in \cite{Bell04} is that the fastest
instability, induced by CR current streaming ahead the shock front,
has a growth rate $\Gamma_{\rm g} \propto n_0^{1/2} V_{\rm sh}^3$
where $n_{\rm 0}$ and $V_{\rm sh}$ are the ambient gas density and the
shock velocity, respectively. Hence, the largest magnetic field
fluctuation growth rates produced by energetic particles at an energy
$E$ are obtained in dense environments pervaded by fast shocks. Some
authors \cite{Schure13, Marcowith14, Cardillo15} have therefore
pointed to the earliest stages of SN evolution (within months to years
of explosion) as possible PeVatron accelerators.

Core-collapse SNe arise from massive stars, which lose considerable
mass during their lifetime. The SNe thus expand into the wind region
formed by the progenitor star. For constant wind mass-loss parameters
(mass-loss rate and wind velocity) the density of the region drops as
r$^{-2}$, and thus is maximum close in to the star. One possibility
therefore would be to search for gamma-ray emission at a very early
expansion stage, when the forward shock is interacting with the very
dense circumstellar medium (CSM). GeV gamma-rays and neutrinos appear
to be the best opportunities to test particle acceleration and CR
production in SNe \cite{Murase14}. However GeV photons associated with
interaction-powered SNe have not been detected in a Fermi-LAT data
search of a sample of 147 SNe of type IIn and Ibn
\cite{Ackermann15}. A search of 45 super-luminous supernovae (SLSNe)
with the Fermi-LAT telescope \cite{Renaultetal17} also did not find
any excess $\gamma$-rays at the SLSN positions.  A recent study of 10
archival SNe \cite{hesssne}, observed with the H.E.S.S. Observatory
\cite{Hess15} within a year after explosion, found no significant
evidence of TeV gamma-ray emission from any of the young SNe.

Our goal in this paper is a proper evaluation of the gamma-ray
emission during the early phase of blast wave expansion. Following the
approach adopted in \cite{Dwarkadas13} we derive a general formalism
including SN dynamics and wind properties, which can be applied to any
SN type where self-similar solutions \cite{Chevalier82} are
applicable. We assume that the self-similar solutions are applicable
even when some of the SN energy is expended in accelerating
particles. This is a reasonable assumption provided that the CR
pressure does not exceed $\sim$ 10\% of the gas pressure
\cite{Chevalier83,Kang10}.

The main hypothesis driving our study is that CR-driven plasma
instabilities lead to the magnetic fields deduced from radio
monitoring of SNe (see {\cite{Marcowith14} and \cite{Bykov18}}).
Starting from this assumption we adapt the theory of diffusive shock
acceleration \cite{Drury83, Berezhko99} to the case of fast moving
forward shocks expanding into the CSM produced by the wind of the SN
progenitor star. Within the adopted formalism we discuss the different
instabilities that may lead to MFA, and test CR acceleration
efficiency at the forward shock for core-collapse SNe. We also include
an accurate treatment of the evolution of the CR maximum energy with
time.

The results obtained herein are quite general, and applicable to any
core-collapse SN whose ejecta density profile and surrounding medium
density profile can be described by a power law. Specific calculations
are made for the case of SN 1993J.

\section{Shock dynamics}
\label{S:DYN}
The shock radius and velocity are assumed to evolve as a power-law
with time. The initial time after the SN outburst is $t_0$ and the
corresponding shock radius is $R_0$. We have: 
\begin{equation}
\label{Eq:RSH}
R_{\rm sh}(t)=R_0 \times \left({t \over t_0}\right)^{m} \ ,\,\,\,\, V_{\rm sh}(t)={R_0 m \over t_0} \times \left({t \over t_0}\right)^{m-1} 
\end{equation}

We note $V_0 =R_0 m / t_0$. If the ejecta density of the SN
${\rho}_{ej} = A t^{-3} {\rm v}^{-k}$, and the surrounding medium
density ${\rho}_{cs} = C\,r^{-s}$, then a self-similar solution for
the shock evolution gives \cite{Chevalier94}:

\begin{equation}
\label{eq:racss}
R_{\rm sh}(t)= \beta \,\left({\frac{\alpha A}{C}}\right)^{1/(k-s)} \; t^{(k-3)/(k-s)} \ ,
\end{equation}

\section{Wind density profile}
The wind mass density scales as a power-law with an index $s$ which
depends on the mass-loss history of the progenitor. For a steady wind
(constant mass-loss rate and wind velocity) $s=2$. The mass density
experienced by the forward shock at a time $t$ is, using
Eq.(\ref{Eq:RSH}), 

\begin{equation}
\label{Eq:RHO} 
\rho_{\rm CSM}(t)=\rho_0~\left({R_{\rm sh}(t)\over R_0}\right)^{-s}= \rho_0~\left({t \over t_0}\right)^{-ms} \ , 
\end{equation}

\begin{equation}
\label{Eq:DEN}
\rho_0 ={\dot{M}(R_0) \over 4\pi V_{\rm w}(R_0) R_{\rm 0}^2} \simeq 1.3m_{\rm p} n_{\rm H, 0} \ , 
\end{equation}

\noindent
where the factor 1.3 accounts for the presence of a medium containing
90\% H and 10\% He, and $m_{\rm p}$ and $n_{\rm H,0}$ are the proton
mass and hydrogen density at $t_0$. We have for the CSM density
at $R_0$
\[\rho_{0} \simeq \left[{5.0 \times 10^{13}\over R_{\rm{0}}^2}~\rm{g/cm^{3}}\right] ~\dot{M}_{-5}(R_0)~V_{\rm w,10}(R_0)^{-1} \ ,\]
where, the shock radius at $t_0$ is expressed in cm, the progenitor
mass-loss rate $ \dot{M}$ is expressed in units of $10^{-5} M_{\odot}
\rm{~yr}^{-1}$ and the wind asymptotic speed $V_{\rm w}$ is in units
of 10 km/s. The mass-loss rate is derived at a fixed radius $R_{\rm
  ref} = 10^{15}$ cm (see \cite{Fransson96}).  The mass-loss rate at
$R_0$ is by definition given by $\dot{M}(R_0)=\dot{M}(R_{\rm
  ref})\left(R_{\rm ref}/R_0\right)^{2-s}$. We consider the wind
velocity to be constant with the radius.

\section{Magnetic field strength}
The magnetic field strength at the stellar surface obtained by a
balance between magnetic field energy density and wind kinetic energy
density is:
\[B_{\rm eq,0} \simeq \left[{2.5 \times 10^{13} \over R_{\rm 0}}~\rm{G}\right]~\dot{M}_{-5}^{1/2}V_{\rm w,10}^{1/2} \ .  \]

We assume a CSM magnetic field strength proportional to $B_{\rm eq}$
with
\begin{equation}
\label{Eq:BWI} 
B_{\rm  w}(t) \simeq \varpi B_{\rm eq,0} \left({t  \over t_0}\right)^{-ms\over 2}
\end{equation}

\noindent
where we assume the ratio $\varpi= B_{\rm w}(t_0)/B_{\rm eq,0}$ to be in the range 0.1--10. The time dependence arises from the radial dependence of the wind density as mentioned in section \ref{S:DYN}.
 
In Eq. (\ref{Eq:BWI}) as soon as $R(t) \gg R_\star$, the wind magnetic
field scales as $1/R_{\rm sh}$ which is expected in case of a toroidal
geometry. The ambient Alfv\'en velocity $V_{\rm A,CSM} = B_{\rm
  W}/\sqrt{4\pi \rho_{\rm CSM}} = \varpi V_{\rm w}$ and the CSM
magnetization ${\cal M}= (V_{\rm A,CSM}/c)^2 = {\cal M}\simeq
\left[1.1~10^{-9}\right] \varpi^2 V_{\rm w,10}^2 $.  Considering
$\varpi$ to be in the range 0.1-10 we always obtain ${\cal M} \ll 1$
whatever the type of progenitor.

\section{SN 1993J}
Supernova 1993J, at a distance of 3.63 Mpc \cite{Freedman01}, became
the optically brightest SN in the northern hemisphere. It resulted
from the explosion of a massive star in a binary system with a
progenitor mass ranging in the interval 13-20 $M_{\odot}$
\cite{Maund04}. The star then evolved into a red super-giant (RSG)
phase with a mass loss rate of $\sim 10^{-6}$ to $10^{-5} M_{\odot}
\rm{~yr}^{-1}$ and a slow wind $V_{w} \sim 10~\rm{km/s}$
\cite{Tatischeff09}.

\section{Acceleration models}
\label{S:GEN}
We adopt a model for particle acceleration at collisionless shocks
based on the theory of DSA \cite{Drury83}. The highest energy CRs have an
upstream diffusion coefficient $\kappa_{\rm u}$ which fixes the length
scale of the CR precursor $\ell_{\rm u}= \kappa_{\rm u}/V_{\rm
  sh}$. The timescale to advect the frozen CR-magnetized fluid to the
shock front is 
$T_{\rm adv,u}= {\kappa_{\rm u} \over V_{\rm sh}^2}  .$
CRs at energies close to $E_{\rm max}$ stream ahead of the shock
and simultaneously generate electromagnetic fluctuations. The upstream
diffusion coefficient at these energies can be expressed with respect
to the diffusion coefficients parallel and perpendicular to the
background wind magnetic field. This coefficient depends on
\cite{Jokipii87}: $\eta$, the ratio of the parallel CR mean free path
to CR Larmor radius $R_{\rm L}$, and $\theta_{\rm B}$ the magnetic
field obliquity. The parallel diffusion coefficient
$\kappa_{\parallel} = \eta R_{\rm L} v/3$, where $v$ is the particle
speed.  $\eta=1$ corresponds to the Bohm diffusion limit. In parallel
shocks ($\theta_{\rm B}$ =0) $\kappa_{\rm u} = \kappa_{\parallel}$
while in perpendicular shocks ($\theta_{\rm B}= \pi/2$) it matches the
perpendicular diffusion coefficient, i.e. $\kappa_{\rm u} =
\kappa_{\perp}$. Without considering magnetic field line wandering in
the wind turbulent medium we have $
\kappa_{\perp}=\kappa_{\parallel}/(1+\eta^2)$. Hence, if $\eta \gg 1$
diffusion is suppressed in the perpendicular shock case. If the
magnetic field in the wind is purely toroidal and weakly turbulent the
advection timescale $T_{\rm adv, u}$ drops. If the wind medium has
some level of turbulence then we can expect to have a diffusion
coefficient close to Bohm ($\eta \sim 1$), and a non negligible
portion of the shock in the parallel configuration.

We define as model P and model T the two extreme configurations
described above. In model P the wind magnetic field is assumed to be
parallel. The advection time in this case is 

\beq
\label{Eq:ADVP} 
T_{\rm adv,u,P}\simeq {\eta_{\rm P} R_{\rm L} v \over 3 V_{\rm sh}^2} \ .  
\eeq 

Accounting for some turbulence in the wind medium, we include a
contribution due to perturbations in the wind magnetic field, $\delta
B_{\rm u}$, which is assumed to be in equipartition with the mean
field strength $B_{\rm w,0}$: $B_{\rm w}^2 = \delta B_{\rm u}^2 +
B_{\rm w,0}^2$. This turbulence is assumed to be injected at large
wind scales, typically the wind termination shock radius, and $\delta
B_{\rm u} \simeq B_{\rm w,0}$ at the highest CR energies.

Using Eq. (\ref{Eq:BWI}) for the wind mean magnetic field, and the
proton Larmor radius $R_{\rm L} \simeq E/ e B_{\rm w}$ for a 1 PeV
particle as $R_{\rm L} \simeq 3.3~10^{12}E_{PeV} B_{\rm
  w,G}^{-1}~\rm{cm}$ \, we find an advection time in seconds

\be
\label{Eq:TAD} 
T_{\rm adv,u,P} \simeq \left[{1.3~10^9 \eta_{\rm P} R_{\rm 0,cm} \over V_{\rm 0, cm/s}^2 \varpi}~\rm{s} \right] \times {E_{\rm PeV} \over \dot{M}_{-5}^{1/2}  V_{w,10}^{1/2}}~\left({t \over t_0}\right)^{2(1-m)+m{s\over2}} \ .  
\ee

In model T the wind mean magnetic field is assumed to be toroidal and weakly perturbed
with fluctuations of strength $\delta B_{\rm u,w} < B_{\rm w, 0} \simeq B_{\rm w}$.  The advection time is in this case
\beq
\label{Eq:ADVT}
T_{\rm adv,u,T} \simeq {R_{\rm L} v \over 3 \eta_{\rm T} V_{\rm sh}^2}\ . 
\eeq 

For the parameters adopted for SN 1993J we have $T_{\rm adv,u,T}
\simeq (0.24~\rm{day})\times (1/\eta_{\rm T}\varpi) E_{\rm PeV} t_{\rm
  d}^{1.17}$ and $T_{\rm adv,u,P} \simeq (0.24~\rm{day})\times
(\eta_{\rm P}/\varpi) E_{\rm PeV} t_{\rm d}^{1.17}$, where $t_{\rm d}$
is the time in days after the SN explosion. \\

We can deduce the acceleration timescale from the above estimates \beq
\label{Eq:TAC} 
T_{\rm acc, P}=g(r) T_{\rm adv,u, P}  = g(r) {\kappa_u \over V_{\rm sh}^2}
\eeq
\noi
where $g(r)=3r/(r-1) \times (1+ \kappa_d r/\kappa_u)$ depends on the
shock compression ratio $r$ and on the ratio of the downstream to
upstream diffusion coefficients. The ratio $\kappa_{\rm d}/\kappa_{\rm
  u}$ depends on the magnetic field
obliquity and on the shock compression ratio $r$. We have
$\kappa_{\rm d}/\kappa_{\rm u} = r_{\rm B}^{-1}$ with $r_{\rm B} = B_{\rm d}/B_{\rm u}$ is the ratio of magnetic fields in the postshock region and in the wind and $g(r)=3r/(r-1)\times (1+r/r_{\rm B})$. In the model P, we have $r_{\rm B} \simeq 1$ and $g(r)=3r(r+1)/(r-1)$. In the model T the magnetic field is weakly perturbed and perpendicular to the shock normal and $\kappa_{\rm d}/\kappa_{\rm u} = r^{-1}$ and $g(r)=6r/(r-1)$. \\

\subsection{Magnetic Field Amplification} Various CR driven instabilities may operate at the
SN forward shock, generating magnetic field fluctuations necessary for
the DSA process to operate at a high efficiency. These include (1)
Bell non-resonant streaming instability \cite{Bell04}: the streaming
of CRs ahead of the shock front induces a return current in the
background plasma, which triggers magnetic fluctuations at scales
$\ell \ll R_{\rm L}$, where $R_{\rm L}$ is the Larmor radius of the
CRs producing the current. This instability is non-resonant and can be
treated using a modified MHD model \cite{Bell04, Bell05, Pelletier06}.
(2) Resonant streaming instability \cite{Amato09}: The streaming of
CRs faster than the local Alfv\'en speed is known to produce
long-wavelength modes at scales $\ell \sim R_{\rm L}$. (3)
Filamentation instability \cite{Reville12}: Cosmic rays form
filamentary structures in the precursors of supernova remnant shocks
due to their self-generated magnetic fields, which results in the
growth of a long-wavelength instability. (4) Long oblique modes
\cite{Bykov11}: The presence of turbulence at scales shorter than the
CR gyroradius enhances the growth of modes with scales longer than the
gyroradius for particular polarizations. Complete details can be found
in \cite{Marcowithetal18}.

\begin{figure}
\centering
\includegraphics[width=8cm]{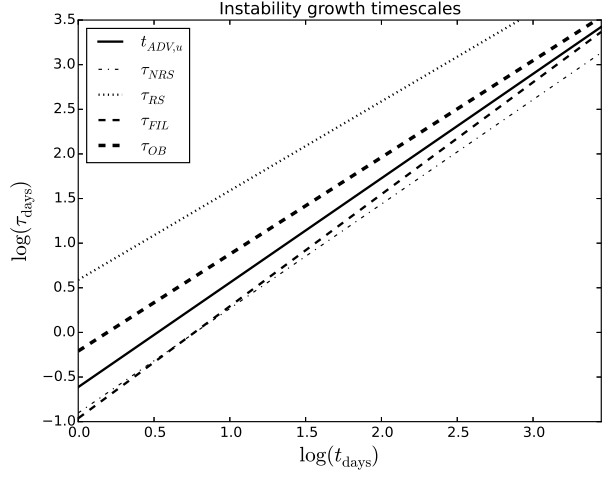}
\caption{Main instability growth timescales as a function of the time in days for the fiducial case SN 1993. We have assumed $\eta=\varpi=1$, $E=1$ PeV, $\phi=14$, $\xi_{\rm CR}=0.05$.}
\label{F:WG}
\end{figure}

Figure \ref{F:WG} plots the advection time and the different growth
timescales for model P for particles with energies of 1 PeV, for the
case of SN 1993J. At all times non-resonant modes can grow. Large
scale modes can be produced by the filamentation instability. The
oblique mode instability and the resonant streaming instability have
growth timescales larger by factors of $\sim 2.5$ and $\sim 15$
compared to the advection time, and can not grow for this set of
parameters.  However, these timescales drop more rapidly with time and
at some stage can become shorter than the advection time, competing
with the filamentation instability to produce long-wavelength
perturbations.\\

\section{Maximum cosmic ray energies}\label{S:MAX}
\begin{figure}
\centering
\includegraphics[width=9cm]{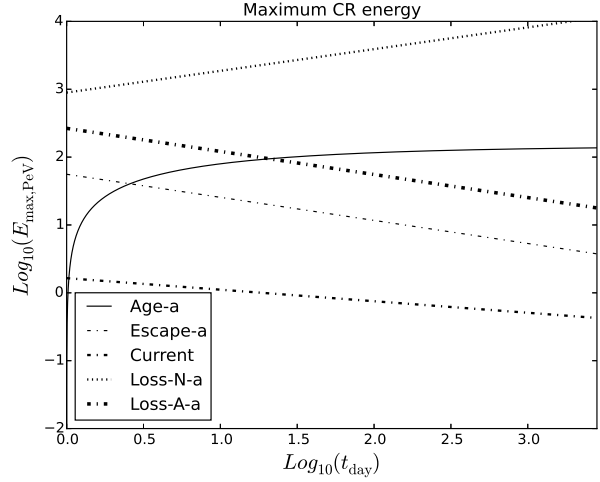}
\caption{Maximum CR energy limits in PeV units for the model P as a
  function of time after shock breakout for SN 1993J. The background
  field has been amplified up to $B_{\rm sat, NRS}$. The dotted line
  plots $E_{\rm max, nuc}(t)$, the large dot-dashed line plots $E_{\rm
    max, adi}(t)$, the intermediate dot-dashed line plots $E_{\rm max,
    cur}(t)$, the small dot-dashed line plot $E_{\rm max,esc}(t)$, the
  solid line plots $E_{\rm max,age}(t)$ We use: $\varpi=1$, $\eta=1$,
  ${\cal N}=5$, $\phi=14$, $\bar{\sigma}_{\rm pp}$=1.87.}
\label{F:EMA}
\end{figure}

The maximum CR (hadronic) energy (Figure \ref{F:EMA}) is fixed by five
different processes: (1) Age limitation, (2) Finite spatial extent of
the shock, (3) Generated current limitation, (4) Nuclear interaction
losses, and (5) Adiabatic losses.

\section{Discussion} \label{S:DIS}
We have shown that SNe can produce particles up to multi-PeV energies
via the combination of fast shocks, a high density CSM produced by
stellar winds, and low wind magnetization.  Assuming that the
background magnetic field has a turbulent component, instabilities
driven by the acceleration process can grow over intra-day
timescales. This model is applied to SN 1993J.\\

{\bf Acknowledgements:} VVD is supported by NASA ADAP grant
NNX14AR63G, and by the FACCTS program. This work is supported by the
ANR-14-CE33-0019 MACH project.

\bibliographystyle{JHEP}
\bibliography{Marco} 

\end{document}